%% file: paper.tex
\documentclass[conference]{IEEEtran}

\usepackage[utf8]{inputenc}
\usepackage[greek,english]{babel}
\usepackage{subcaption}
\usepackage{graphics}
\usepackage{graphicx}
\usepackage{cite}
\usepackage{xcolor}
\usepackage[]{algorithm2e}
\usepackage{listings}
\usepackage{float}
\usepackage[bookmarks=true,%
            bookmarksnumbered=true,%
            colorlinks=true,%
            linkcolor=linkcol,%
            citecolor=citecol,%
            urlcolor=urlcol,%
            hypertexnames=true,%
            pdfpagelabels]%
      {hyperref}
   \hypersetup{ pdfauthor = {Jeremiah Onaolapo},
                pdftitle = {},
                pdfkeywords = {},
                pdfcreator = {},
                pdfproducer = {}
              }

   \definecolor{linkcol}{rgb}{0,0,0.5}
   \definecolor{citecol}{rgb}{0,0.5,0.3}
   \definecolor{urlcol}{rgb}{0.3,0,0}
\usepackage{url}
\usepackage{todonotes}
\usepackage{amsfonts}
\usepackage{amstext}
\usepackage{fixltx2e}
\usepackage{flushend}
\usepackage{authblk}
\usepackage{multirow}
\usepackage{hyperref}
\usepackage[hyphenbreaks]{breakurl}
\usepackage{url}
\usepackage[normalem]{ulem}

\newcommand{\desc}[1]{\noindent\textbf{#1.~}}
\newcommand{\hypl}[1]{Hypothesis #1\xspace}
\newcommand{\hyp}[1]{$H_{#1}$\xspace}

\title{Email Babel: Does Language Affect Criminal Activity in Compromised Webmail Accounts?}

\author{Emeric Bernard-Jones, Jeremiah Onaolapo, and Gianluca Stringhini\\
  University College London\\
  \texttt{emeric.bernard-jones.15@ucl.ac.uk} \\
  \texttt{\{j.onaolapo,g.stringhini\}@cs.ucl.ac.uk}
}

\begin{document}

\maketitle
\thispagestyle{empty}

\input{vars}

\begin{abstract}

We set out to understand the effects of differing language on the ability of cybercriminals to navigate webmail accounts and locate sensitive information in them. To this end, we configured \numOfAccounts{} Gmail honeypot accounts with English, Romanian, and Greek language settings. We populated the accounts with email messages in those languages by subscribing them to selected online newsletters. We hid email messages about fake bank accounts in fifteen of the accounts to mimic real-world webmail users that sometimes store sensitive information in their accounts. We then leaked credentials to the honey accounts via paste sites on the Surface Web and the Dark Web, and collected data for \durationOfExpt. 
Our statistical analyses on the data show that cybercriminals are more likely to discover sensitive information (bank account information) in the Greek accounts than the remaining accounts, contrary to the expectation that Greek ought to constitute a barrier to the understanding of non-Greek visitors to the Greek accounts. We also extracted the important words among the emails that cybercriminals accessed (as an approximation of the keywords that they searched for within the honey accounts), and found that financial terms featured among the top words. In summary, we show that language plays a significant role in the ability of cybercriminals to access sensitive information hidden in compromised webmail accounts. 


\end{abstract}

\section{Introduction}

Online accounts provide many useful functionalities but also expose users to various risks, including information theft. For instance, we send emails, edit online documents, and network with colleagues via online accounts. Consequently, these accounts not only provide these capabilities, but also often become repositories of sensitive information, such as passwords and financial information. Webmail accounts are particularly ``susceptible'' to this, since they store private information by design. This makes them attractive to miscreants that seek to make a fortune from the content of such accounts.  

Data breaches and unauthorised account accesses are commonplace nowadays, usually at high financial and reputation costs to victims and online service providers alike~\cite{anderson2013:cost}.
Cybercriminals usually compromise online accounts by performing social engineering or phishing attacks on victims~\cite{dhamija2006phishing}. Other ways by which cybercriminals obtain credentials and compromise online accounts include database breaches,\footnote{\url{http://krebsonsecurity.com/2014/05/the-target-breach-by-the-numbers/}} information-stealing malware~\cite{stone09:torpig}, and network attacks.\footnote{\url{http://codebutler.com/firesheep}, \url{http://crypto.stanford.edu/ssl-mitm}} 

After obtaining the credentials of online accounts, cybercriminals usually assess the value of the accounts by evaluating the content of the compromised accounts and searching for sensitive information~\cite{bursztein2014handcrafted}. Depending on the perceived value of the accounts, the miscreants then sell the account credentials on the underground black market~\cite{stone2011underground}, or use them privately. In some  cases, the cybercriminals carry out further attacks against the owners of such accounts, for instance by mounting blackmail attacks against them, as seen in the Ashley Madison online dating website scandal.\footnote{\url{https://blog.kaspersky.co.uk/cheating-website-hacked/}} In other cases, the compromised accounts are used to attack other online users, for instance sending spam messages to the contacts of the account owner~\cite{stone2011underground}. 

Existing literature on the use of compromised online accounts by cybercriminals is sparse. This is primarily because it is difficult to collect data on compromised accounts without being in control of a large online service. Bursztein et al. studied Gmail accounts that were compromised via phishing attacks, to understand the modes of operation of cybercriminals that gained illegitimate access to the accounts~\cite{bursztein2014handcrafted}. Similarly, Onaolapo et al. studied the modus operandi of miscreants accessing Gmail accounts leaked through multiple outlets~\cite{onaolapo2016gmail}. Lazarov et al. investigated the activity of miscreants on leaked online spreadsheets~\cite{lazarov2016honey}. 

Online accounts often allow users to customise their accounts in various ways, for instance, through language localisation. This question then comes to mind -- how do cybercriminals behave when they encounter accounts in a different locale or language? How will this affect their activity? To the best of our knowledge, there is limited existing research on this theme. 
To close this research gap, we studied the impact of differences in account language on the activity of miscreants that connect to compromised Gmail accounts. 

To this end, we employed the infrastructure and methodology proposed in our previous paper~\cite{onaolapo2016gmail}. Hence, we created and instrumented thirty Gmail accounts. We populated them with email messages in three different languages, namely English, Greek, and Romanian. We seeded fifteen of the accounts with fake bank details containing keywords that are known to be appealing to cybercriminals. We then leaked credentials to the accounts through paste sites in the Surface and Dark Webs, following the approach employed in previous work~\cite{onaolapo2016gmail}.  
We recorded accesses and activity in the accounts and carried out statistical tests on the collected data. 

We found that cybercriminals are more likely to discover the fake bank account details hidden in the Greek accounts than the remaining accounts. This is contrary to the expectation that Greek ought to constitute a barrier to the understanding of non-Greek visitors to the Greek accounts. Previous work shows that cybercriminals typically assess the value of stolen accounts by searching for valuable information in them~\cite{bursztein2014handcrafted,onaolapo2016gmail}. Thus, we postulate that the cybercriminals possibly used online language translation tools to translate financial terms to Greek prior to searching the Greek accounts for such keywords. This would also explain the amount of time that they spent accessing the accounts: Greek accounts recorded longer access times than the rest, while English accounts recorded the lowest access times. We present detailed results in Section~\ref{sec:analysis}. 

Using Natural Language Processing (NLP) techniques, we extracted important words from the emails that cybercriminals accessed (as an approximation of the keywords that they searched for within the honey accounts), and found that financial terms featured among the top words. This is interesting because some of the sensitive words that we seeded the honey accounts with also showed up among those important words. This indicates that the cybercriminals paid particular attention to those sensitive emails. 

In summary, we found that language indeed affects the ability of cybercriminals to locate sensitive information in the honey accounts. Our statistical tests show that there is a significant relationship between language and criminal activity in webmail accounts. We also corroborate previous findings that cybercriminals search for financial and other sensitive information in compromised webmail accounts~\cite{bursztein2014handcrafted, onaolapo2016gmail}. 

\desc{Contributions}We provide detailed statistical analyses showing that language differentiation affects the ability of cybercriminals to locate sensitive information in a compromised webmail account. To the best of our knowledge, this is the first study that explores the relationship between language and criminal ability.

\section{Background}
\label{sec:background}

In this section, we discuss the categories of cybercrime, webmail accounts, and the relationship between language and criminal ability. Finally, we present our research questions and hypotheses.
 
\subsection{Categories of cybercrime}
\label{sec:background:cybercrime}

Broadly speaking, cybercrime is a term used to describe a wide variety of instances within which technology is used or involved in the execution of a criminal act~\cite{clough2011cybercrime}. It embodies a variety of criminal activities (for instance, identity theft and fraud), many of which are among the most rapidly advancing crime types in many developed countries~\cite{lopucki2001human}. We discuss the three distinct categories of cybercrime, namely cyber-assisted crimes, cyber-dependent crimes, and cyber-enabled crimes~\cite{wall2003mapping}.

\desc{Cyber-assisted crimes}These are terrestrial crimes, such as burglary or theft, which incorporate the use of digital technologies into the execution of a criminal act~\cite{levi2016implications}. An example of this is when a bicycle thief uses a mapping application to plan a route through the area they already intended to steal from. In cyber-assisted crime, the ``cyber'' element plays a tertiary role in the execution of the crime itself, that is, the crime would likely continue unaffected if the cyber element was removed. This type of cybercrime is not in the scope of this paper, but it is useful to note the extent to which advances in information technology are able to influence terrestrial or physical crime types.

\desc{Cyber-dependent crimes}These are crimes that can be executed without the use of an internet connection, but use technology as a force multiplier to commit terrestrial crimes within a ``cyber-sphere''~\cite{mcguire2013cybercrime}. These crimes often take advantage of the global reach of the Internet, but do not necessarily represent entirely new crime types. A clear example is bank fraud which existed before the Internet but has been greatly facilitated by the growth of the Internet.  

\desc{Cyber-enabled crimes}These represent the ``cybercrime archetype.'' These crimes cannot be committed without the use of an internet connection or computer network, for instance, a Distributed Denial of Service (DDoS) attack~\cite{buscher2012ddos}.

Although debate exists regarding small differences and measurement of these crime types~\cite{furnell2015challenge}, our intent in this paper is not to provide detailed insight into crime types or classifications. In this paper, we focus on cyber-dependent and cyber-enabled crimes, since we study the actions of criminals that access the contents of webmail accounts illegitimately.

\subsection{Gmail accounts}
\label{sec:background:gmail}

Gmail accounts, like many other webmail accounts, allow users to send and receive text/multimedia messages to one another. 
However, beyond sending and receiving email messages, Gmail users can embed scripts in their accounts to automatically carry out other activities, for instance, to remind them about important emails that require attention. We leveraged this functionality to instrument the Gmail accounts that we used in our experiments, by configuring the scripts to send us notifications about changes in the accounts (see Section \ref{sec:methodology:monitoring}).

After authenticating to their accounts, Gmail users can access the email messages that other webmail users sent to them in their \textit{Inbox} folder. While composing email messages in preparation for sending to other webmail users, those email drafts appear in the \textit{Drafts} folder. Similarly, they can access the email messages that they previously sent to others in the \textit{Sent} folder. They can mark emails for later reference by \textit{starring} them. Gmail also provides a search tool for users to enter search terms when looking for emails containing those terms. Finally, Gmail users can change the display language of their Gmail interface so that menu items, options, and text on Gmail pages will be displayed in the selected language. This is particularly useful for non-English Gmail users.  

\subsection{Language and crime}
\label{sec:background:language}

Research suggests that criminal activities are carried out along familiar patterns of behaviour, spatially, by crime type, or by the network of the actors~\cite{brantingham1995criminality}. This therefore suggests that successful criminals rely heavily on a detailed understanding of the processes surrounding the crimes they commit and the areas within which they are committed~\cite{pizarro2007journey}. Thus, we can safely assume that their ability to understand and interpret social cues, their environment, and the behaviour of their victims has a knock-on effect on their ability to commit crime~\cite{brantingham1993nodes}. 

While attempting to study the behavioural patterns of criminals online, connecting to a webmail account and navigating through it can be considered a ``routine activity,'' since these are frequent online actions by legitimate users. Changes in the composition, interface, layout, or language of the webmail account can therefore be considered a barrier to the execution of a crime in the account -- much like a physical barrier (for instance, a fence) may deter terrestrial crime. This forms the thematic basis for our work.

Certain other aspects of criminal theory developed for terrestrial crime types have shown promise in their ability to be adapted to fit cybercrime types~\cite{yar2005novelty}. Even though ideas of locality or geographical nodes from crime pattern theories may need to be replaced with cyber equivalents, certain trends and routine activities online have been successfully attributed to specific online criminals~\cite{spitters2015darkmarket}. 

There is a commonplace ``truism'' when discussing cybercrime: that cybercrime is somehow unrestricted by the same boundaries of time, space, and culture that may hinder traditional crime types~\cite{grabosky2004global}. However, the majority of contentions in previous work were made through logical inferences and assertions. In particular, after exploring previous work in the domains of crime sciences and language, we found very little research exploring the relationship between language and crime. This paper seeks to close that research gap and provide insights into whether the execution of a criminal act is indeed affected by language differences and comprehension or not. To this end, we define our research question and hypotheses as follows.

\desc{Research question}Does language differentiation affect cybercriminal activity?

\desc{\hypl{0} (\hyp{0})}Language differentiation will not have a significant impact on the ability of cybercriminals to locate a sensitive item in a compromised webmail account.

\desc{\hypl{1} (\hyp{1})}Language differentiation will have a significant impact on the ability of cybercriminals to locate a sensitive item in a compromised webmail account.

\section{Methodology} 
\label{sec:methodology}

In this section, we describe the creation, population, and seeding of the honey accounts. We also describe our method of collecting data from the honey accounts.

\subsection{Creating honey accounts} 
\label{sec:methodology:creating}

We created thirty honey accounts on the Gmail service across three languages, namely English (ten accounts), Romanian (ten accounts), and Greek (ten accounts). We chose those languages for linguistic reasons; English because it is an ``international'' language, Romanian because it is the only Latin-based Eastern European language, and Greek because it features a unique alphabet. In order to minimise potential biases in our dataset, we configured the fake personas of the honey accounts such that each linguistic group comprised five men and five women, with birth dates ranging from 1960 to 2000. We did this to make the accounts appear as believable as possible by featuring a diverse persona set.

\begin{figure}[tp]
	\centering
	\fbox{\includegraphics[width=.4\textwidth]{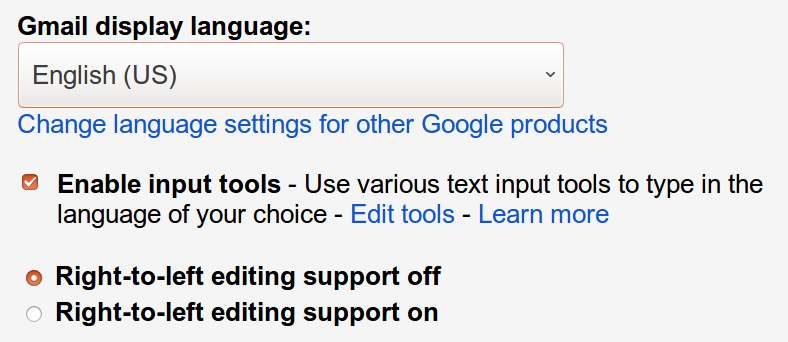}}
	\caption[]{Gmail gives users the option to change the display language of its user interface. In addition to populating the honey accounts with language-specific newsletters, we also changed the display language of each honey account to match its contents. }
	\label{fig:gmail-lang-setting}
\end{figure}

To populate the accounts, we subscribed them to over fifty language-specific newsletters and mailing lists following certain themes we had previously selected. The themes include fashion, law, and gardening, and were picked according to the gender and date of birth of the fake personas we developed for the honey accounts. We also changed the display language of each honey account to match the language of its content. Figure \ref{fig:gmail-lang-setting} shows the Gmail language configuration option that allows this. 

\begin{figure}[tp]
	\centering
	\fbox{\includegraphics[width=.4\textwidth]{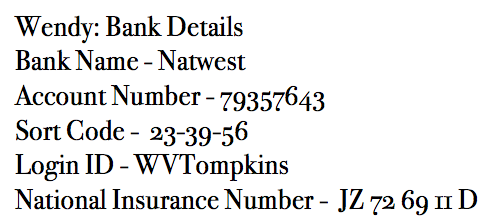}}
	\caption[]{An example of the fake banking details that we hid in the English honey accounts.}
	\label{fig:eng-sortcode}
\end{figure}

\begin{figure*}[tp]
	\centering
	\fbox{\includegraphics[width=.8\textwidth]{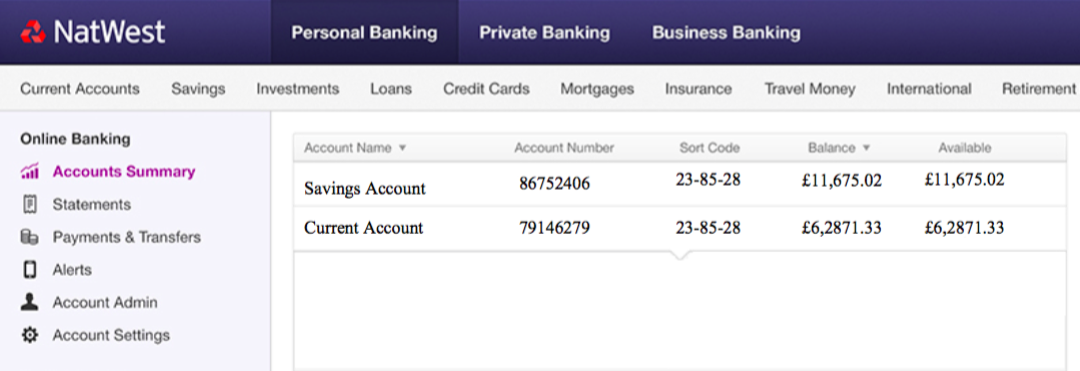}}
	\caption[]{An example screenshot of a fake online banking profile that we hid in the English honey accounts.}
	\label{fig:online-banking}
\end{figure*}

\desc{Sensitive emails} In fifteen out of thirty honey accounts, we hid fake online banking information. The idea was to mimic the behaviour of webmail users that store sensitive information in their accounts. To achieve this, we created screenshots of fake bank account details and online banking pages (see Figures \ref{fig:eng-sortcode} and \ref{fig:online-banking}), and sent emails containing the screenshots to the honey accounts themselves. For instance, for each honey account $h_G$ in the accounts designated to contain sensitive information, we sent the screenshots described earlier from $h_G$ to itself.  We used region-specific bank information while seeding the accounts, for instance, fake Natwest and Santander information for English accounts, fake ING information for Romanian accounts, and fake Alpha Bank profiles for Greek accounts. We did this to ensure that the banks would be instantly recognizable in the countries of the honey account personas. We also included keywords such as ``national insurance number,'', ``sort code,'' and ``account number'' in the sensitive emails. Such keywords have been shown to be attractive to cybercriminals~\cite{bursztein2014handcrafted, onaolapo2016gmail}. Finally, we left the remaining fifteen accounts unseeded as the control experiment.

\begin{figure}[tbp]
	\centering
	\includegraphics[width=.4\textwidth]{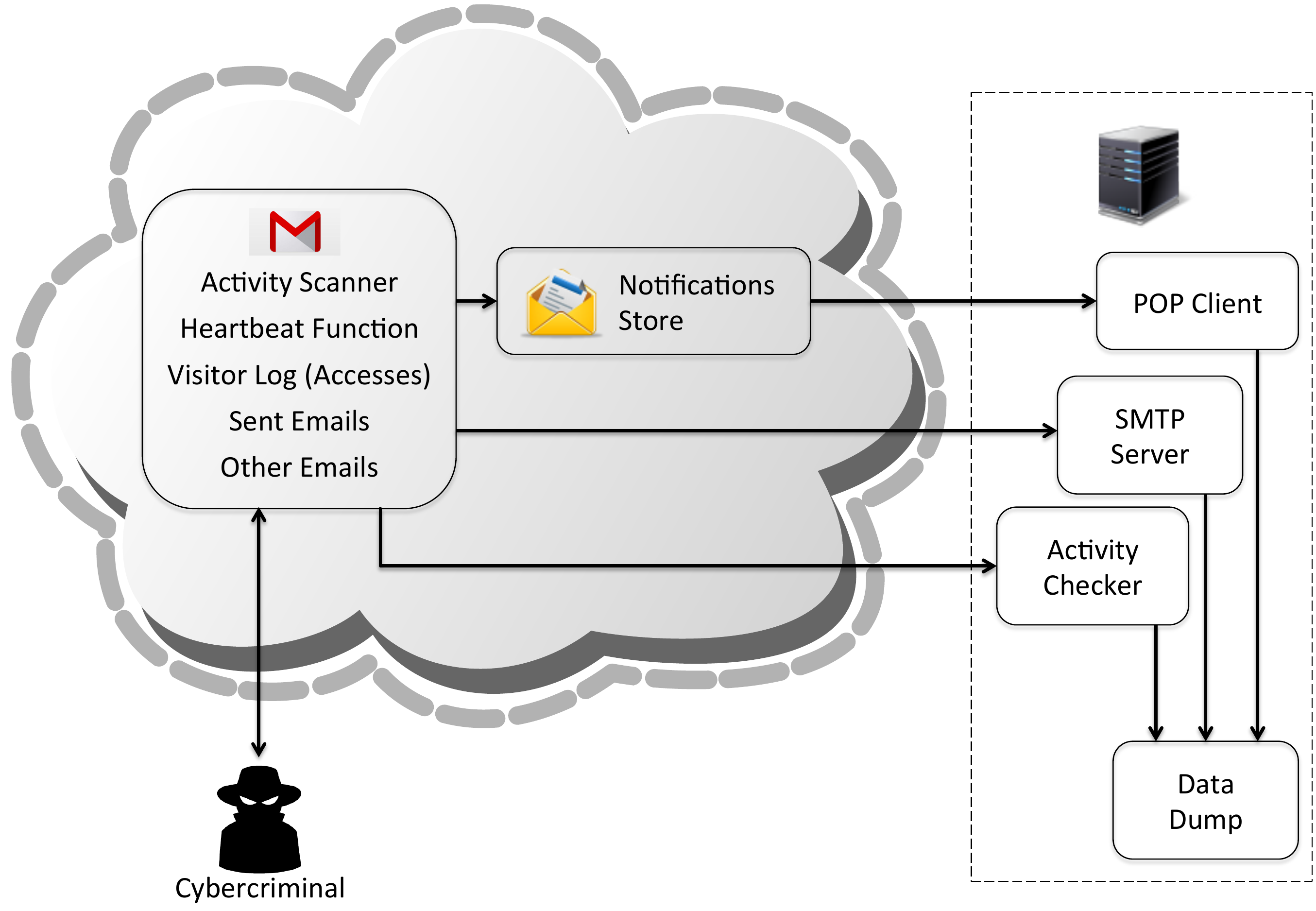}
	\caption[]{Overview of the honeypot system.}
	\label{fig:gmail-sys-overview}
\end{figure}

\subsection{Monitoring honey accounts} 
\label{sec:methodology:monitoring}

To monitor illegitimate activity in the honey accounts, we used the infrastructure presented in our previous paper~\cite{onaolapo2016gmail}. It comprises scripts embedded in the honey accounts, a sinkhole email server, a notification store to receive activity  notifications from honey accounts, an email client to retrieve email messages from the notification store, and some other monitor scripts. Figure \ref{fig:gmail-sys-overview} shows an overview of the monitor infrastructure.  

The system provides us with information about activity in honey accounts, specifically when emails are opened, sent, or starred. It also provides us with information about draft emails created by visitors to the honey accounts. In addition, we receive ``heartbeat'' messages daily from each honey account to notify us about accounts that are active. We cease to receive ``heartbeat'' messages from an account if it has been suspended by Google, or if it was hijacked completely by cybercriminals, that is, if they changed the account's password. Finally, the system provides us with information on accesses to the honey accounts, that is, we receive IP address information, location information, access times, and other details about visitors interacting with the honey accounts. More details about the infrastructure can be found in our previous paper~\cite{onaolapo2016gmail}.

To minimize the risk of abuse, we configured the honey accounts' default send-from addresses to point to an email server which is part of the monitor infrastructure described earlier. Hence, all emails sent from the honey accounts would be delivered to our email server and not to the outside world, since our email server is a sinkhole server (it does not forward emails to the intended destination).

\subsection{Leaking honey accounts} 
\label{sec:methodology:setup} 

After instrumenting the honey accounts, we leaked their credentials via paste sites on the Surface Web and the Dark Web, namely on \texttt{Pastebin}, \texttt{Insertor}, and \texttt{Stronghold}. \texttt{Insertor}, and \texttt{Stronghold} are Dark Web paste sites, accessible only through special software, such as TOR browser. \texttt{Pastebin} is accessible via any common web browser, for instance Firefox, Chrome, or Safari.  
In each leak, we included honey account credentials and messages indicating that the credentials were obtained from hacked accounts. We recorded accesses made to the honey accounts by miscreants and analysed the resulting dataset. Details of our analysis can be found in Section \ref{sec:analysis}.  

\subsection{Threats to validity}

It is important to mention that the monitoring infrastructure we used in this study can only detect if an email was opened, and not necessarily if it was read. For the purpose of this study, we assume that opened emails were also read by the person that opened them. In addition, we currently lack a way to determine the exact words that were searched for in the honey accounts by cybercriminals. Instead, we approximate those search terms by evaluating important words in the emails that were opened by the cybercriminals. We consider this the main threat to the internal validity of this study. To minimize the impact of this threat, we seeded the accounts with email messages containing sensitive content (fake financial information) and hid the emails, such that finding them would require some effort by the cybercriminals. We then focused our analysis on those sensitive emails. In future work, we hope to find a more accurate way to determine search terms in the honey accounts. Another threat to internal validity is that many of the honey accounts were hijacked at least once by cybercriminals, that is, the passwords of such accounts were changed. Recall that we are unable to collect access and activity information from a honey account when that happens. However, it is important to note that we were able to recover some of the accounts and continue the experiments. Finally, we leaked account credentials through paste sites only, therefore, our results may not necessarily reflect what happens when accounts are compromised via other outlets.

\subsection{Ethics}

Due to the sensitive nature of our study, we ensured that the experiments were carried out in an ethical manner. Since the experiments require releasing account credentials to cybercriminals, there is the risk of abuse. We minimized this risk by configuring the honey accounts to send all outgoing emails to an email server under our control, which does not deliver the emails to their intended destinations. Thus, we were able to prevent the accounts from being used to spam other users. Also, we seeded the honey accounts with financial information such as bank accounts and online banking information. To avoid harming anyone, we ensured that all the financial details loaded in the accounts were fake (we generated them randomly). Finally, since our experiments involve deceiving cybercriminals to engage with fake accounts, we obtained ethics approval from our institution.

\section{Data Analysis} 
\label{sec:analysis}

A Gmail account keeps records of each unique access and labels the access with a unique identifier also known as a ``cookie,'' along with other information about the access, such as access time, IP address, and location. We extracted this information from the honey accounts via our honeypot infrastructure (cf. Section \ref{sec:methodology}). We also evaluated the actions corresponding to those accesses (for instance, email opening, sending, starring, or draft creation). In other words, each data unit encapsulates an ``access--action.'' 
 
During our observation period of \durationOfExpt, we observed \numOfAccesses{} data units across 29 honey accounts from \numOfCountries{} countries. We removed 210 of those data units from our dataset (outliers) due to their undue effect on the distribution of the data, bringing the overall total of individual data points to 440. The outliers comprise the data points that emanated from cybercriminals that ran amok in the honey accounts, either reading all emails in the affected accounts or performing a lot of other actions.  
In this section, we present the results of statistical tests and textual analysis on the data. We establish the relationship between language and cybercriminal ability, and also show the keywords that cybercriminals were interested in.

\subsection{Statistical tests}

\begin{table*}[htbp]
\centering
\caption{Chi-squared ($\chi^2$) analysis showing the differences between expected and actual criminal access to a sensitive item.}
\label{tab:chi-sq-sensitive}
\begin{tabular}{l|l|l|l|l|l|l|}
\cline{2-7}
 & \multicolumn{2}{c|}{Not Sensitive} & \multicolumn{2}{c|}{Sensitive} & \multicolumn{2}{c|}{Total} \\ \hline
\multicolumn{1}{|l|}{Language} & Frequency & \begin{tabular}[c]{@{}l@{}}Expected \\ Frequency\end{tabular} & Frequency & \begin{tabular}[c]{@{}l@{}}Expected \\ Frequency\end{tabular} & Frequency & \begin{tabular}[c]{@{}l@{}}Expected \\ Frequency\end{tabular} \\ \hline
\multicolumn{1}{|l|}{English} & 189 & 177.9 & 29 & 40.1 & 218 & 218 \\ \hline
\multicolumn{1}{|l|}{Greek} & 80 & 93.8 & 35 & 21.2 & 115 & 115 \\ \hline
\multicolumn{1}{|l|}{Romanian} & 90 & 87.3 & 17 & 19.7 & 107 & 107 \\ \hline
\multicolumn{1}{|l|}{Total} & 359 & 359 & 81 & 81 & 440 & 440 \\ \hline
\end{tabular}
\end{table*}

\begin{table*}[htbp]
\centering
\caption{Logistic regression assessing the relationship between language and criminal ability to locate a sensitive item.}
\label{tab:regression-language}
\begin{tabular}{|l|l|l|l|l|l|l|}
\hline
Sensitive & Odds Ratio & Std. Err. & $z$ & $ P>|z| $ & \multicolumn{2}{l|}{95\% Confidence Interval} \\ \hline
Lang--Eng & 0.8123249 & 0.2690604 & -0.63 & 0.530 & 0.4244168 & 1.554773 \\ \hline
Lang--Gre & 2.316176 & 0.7716938 & 2.52 & 0.012 & 1.205513 & 4.450116 \\ \hline
Cons & 0.1888889 & 0.049952 & -6.30 & 0.000 & 0.1124876 & 0.3171816 \\ \hline
\end{tabular}
\end{table*}

We coded the collected data into nine variables, namely {\em account name, language of account, email-subject, activity performed, sensitivity of item accessed, IP address, location, country,} and {\em duration of access}. To determine whether a relationship exists between language and cybercriminal functionality, we ran a chi-squared ($\chi^2$) test~\cite{pearson1900chi} to assess any possible associations between discrete languages variables (Greek, Romanian, and English), and the ability of the cybercriminal to access a sensitive item. 

The Pearson $\chi^2$ test (see Table \ref{tab:chi-sq-sensitive}) shows that there is indeed a significant association between language and the ability of a cybercriminal to locate sensitive items within an email account ($\chi^2(2) = 15.3097$, $p < 0.001$). Due to the risk of inflation, we also generated a Cramer's $V$ statistic~\cite{cramer1946vstat} to reveal further information about the strength of the association. This confirmed that there was a weak, yet significant, association between language and cybercriminal ability ($V = 0.1865$). However, it must be noted that $\chi^2$ tables are relatively unable to provide more substantive information regarding the interactions between the variables or the fit of the model implemented. Thus, we carried out logistic regression to further explore if a substantive relationship exists among the three language variables and criminal activity (see Table \ref{tab:regression-language}). We found that the language variables, in combination, significantly affected the ability of a cybercriminal to find a sensitive item ($\chi^2(3) = 19.77$, $p < 0.001$), with the model accurately predicting 81.59\% of criminal action. Note that we dropped the Romanian data points from the analysis in name due to collinearity, and we henceforth refer to them as {\em Cons} in subsequent analyses (that is, in Tables \ref{tab:regression-language}, \ref{tab:regression-time}, and \ref{tab:regression-mean-time}). 

Further analysis revealed a significant positive relationship between the ability to locate a sensitive item and Greek language sets ($z = 2.52$, $p < 0.01$) with an odds ratio of 2.316176, meaning that accounts established in Greek are more than twice as likely to have a sensitive item accessed than either of the other language sets. English language, as a variable, was not significant, ($z = -0.63$, $p=0.530$) with an odds ratio of 0.8123249. This means that an account being constructed in English actually lessens the chance of a miscreant accessing a sensitive item in it. We obtained similar results for the Romanian account set, which was significant ($z = -6.30$, $p < 0.01$), with an odds ratio of 0.1888889. This indicates that there is a significant negative relationship between emails written in Romanian and the ability of a criminal to locate a sensitive item in them.

\begin{table*}[htbp]
\centering
\caption{Logistic regression including access durations.} 
\label{tab:regression-time}
\begin{tabular}{|l|l|l|l|l|l|l|}
\hline
Sensitive & Odds Ratio & Std. Err. & z & $ P > | z | $ & \multicolumn{2}{l|}{95\% Confidence Interval} \\ \hline
Lang--Eng & 0.8618208 & 0.2878145 & -0.45 & 0.656 & 0.4478668 & 1.658384 \\ \hline
Lang--Gre & 2.345972 & 0.7901058 & 2.53 & 0.011 & 1.212396 & 4.539428 \\ \hline
Access & 1.008337 & 0.0038651 & 2.17 & 0.030 & 1.00079 & 1.015941 \\ \hline
Cons & 0.1589789 & 0.0445502 & -6.56 & 0.000 & 0.091793 & 0.27534 \\ \hline
\end{tabular}
\end{table*}

We further introduced {\em access duration} as a variable into logistic regression (see Table \ref{tab:regression-time}). This is because we observed that the mean of the average access rates for accounts across languages varied; Greek accounts had the highest access times on average while the English accounts had the lowest. This might indicate further activity such as content translation to facilitate navigation through the honey accounts. Logistic regression with access duration included among the discrete language variables was significant, accurately predicting 82.05\% of criminal activity, and accounting for a small level of variance within the model ($z = 2.17$, $p < 0.01$). The access time variable also had a slight positive effect on the significance levels represented by the Greek and English variables, with an English odds ratio of 0.8618208 ($z = 0.45$, $p = 0.656$) and a Greek odds ratio of 2.345972 ($z = 2.53$, $p < 0.01$). However, the Romanian variable suffered a corresponding decrease (odds ratio 0.1589789) while still remaining significant ($z = -6.56$, $p < 0.01$).

\begin{table*}[htbp]
\centering
\caption{Logistic regression with mean centralised access durations.}
\label{tab:regression-mean-time}
\begin{tabular}{|l|l|l|l|l|l|l|}
\hline
\multicolumn{1}{|c|}{Sensitive} & \multicolumn{1}{c|}{Odds Ratio} & \multicolumn{1}{c|}{Std. Err.} & \multicolumn{1}{c|}{z} & \multicolumn{1}{c|}{$ P > | z | $} & \multicolumn{2}{c|}{95\% confidence Interval} \\ \hline
Lang--Eng & 0.8618208 & 0.2878145 & -0.45 & 0.656 & 0.4478668 & 1.658384 \\ \hline
Lang--Gre & 2.345972 & 0.7901058 & 2.53 & 0.011 & 1.212396 & 4.539428 \\ \hline
C--Access & 1.008337 & 0.0038651 & 2.17 & 0.030 & 1.00079 & 1.015941 \\ \hline
Cons & 0.1804734 & 0.0482639 & -6.40 & 0.000 & 0.1068506 & 0.3048244 \\ \hline
\end{tabular}
\end{table*}

To re-affirm our findings, we mean-centered the access duration values before running the model again to ensure that the logistic model was not centering the access duration values at an intercept with a value of 0, but rather a value integral to the rest of the model (see Table \ref{tab:regression-mean-time}). Mean-centering had no effect on the fit of the model overall, other than marginally improving the significance of the Romanian language variable ($z = -6.40$, $p < 0.01$), resulting in the final odds ratio of 0.1084737.

Since these results clearly demonstrate that there is a significant relationship between language and cybercriminal ability, we reject our null hypothesis \hyp{0}. In the next section, we present our findings on the items that cybercriminals searched for in the honey accounts.

\subsection{Digging for webmail ``gold''}

We wanted to understand the themes and words that cybercriminals search for when they access compromised webmail accounts. Previous research has shown that one of the first steps of cybercriminals after compromising an online account is to assess its value by going through its contents~\cite{bursztein2014handcrafted}. This implies that they run certain search queries to locate email messages of interest to them. However, we did not have access to the search terms in the honey accounts since there is currently no API to retrieve such information from the honey accounts. To overcome this limitation, we approximated the search terms by analysing the opened emails and extracting the important words in them, relative to all the emails in the honey accounts. To achieve this, we used Term Frequency--Inverse Document Frequency (TF-IDF) analysis, following the method outlined in our previous paper~\cite{onaolapo2016gmail}.

For each language set (English, Greek, Romanian), consider $d_R$ as the corpus of all opened emails in the honey accounts of that language, while $d_A$ is the corpus of all emails in the inboxes of those accounts. We removed all words that had less than five characters from the corpus, and also removed signalling and header information, for instance ``charset.'' We obtained tfidf$_R$ and tfidf$_A$  as the resulting vectors of words and their probabilities after performing TF-IDF analysis on the text corpus [$d_R$, $d_A$]. We further computed the vector tfidf$_R - $tfidf$_A$. The idea is that words with higher tfidf$_R - $tfidf$_A$ values have higher importance in the set of emails opened by miscreants, relative to the entire corpus. Thus, such words reveal the themes that the cybercriminals were likely searching for.

\begin{table*}[htbp]
\centering
\caption{TF-IDF results for the English language variant.}
\label{tab:tfidf-english}
\begin{tabular}{|l|l|l|l|l|l|l|l|}
\hline
\multicolumn{1}{|c|}{Searched words} & \multicolumn{1}{c|}{tfidf$_R$} & \multicolumn{1}{c|}{$tfidf_A$} & \multicolumn{1}{c|}{tfidf$_R - $tfidf$_A$} & \multicolumn{1}{c|}{Common words} & \multicolumn{1}{c|}{tfidf$_R$} & \multicolumn{1}{c|}{tfidf$_A$} & \multicolumn{1}{c|}{tfidf$_R - $tfidf$_A$} \\ \hline
written & 0.4371 & 0.04322 & 0.3938 & unsubscribe & 0.109 & 0.1833 & -0.0743 \\ \hline
question & 0.447 & 0.0678 & 0.3796 & click & 0.0953 & 0.1671 & -0.0718 \\ \hline
answer & 0.2283 & 0.0377 & 0.1907 & please & 0.0931 & 0.1597 & -0.0666 \\ \hline
commission & 0.2224 & 0.0386 & 0.1838 & about & 0.0761 & 0.1279 & -0.0518 \\ \hline
union & 0.2273 & 0.0565 & 0.1708 & service & 0.0394 & 0.1248 & -0.0854 \\ \hline
european & 0.2508 & 0.088 & 0.1628 & twitter & 0.0257 & 0.1193 & -0.0936 \\ \hline
source & 0.2267 & 0.0663 & 0.1604 & trump & 0.0399 & 0.1085 & -0.0685 \\ \hline
banking & 0.1599 & 0.0394 & 0.1205 & london & 0.2158 & 0.1017 & -0.1141 \\ \hline
london & 0.2158 & 0.1017 & 0.1141 & contact & 0.0465 & 0.1001 & 0.0536 \\ \hline
investment & 0.0548 & 0.0122 & 0.0425 & health & 0.0717 & 0.0983 & -0.026 \\ \hline
\end{tabular}
\end{table*}

\begin{table*}[htbp]
\centering
\caption{TF-IDF results for the Greek language variant.}
\label{tab:tfidf-greek}
\begin{tabular}{|l|l|l|l|l|l|l|l|}
\hline
Searched Words & tfidf$_R$ & tfidf$_A$ & tfidf$_R - $tfidf$_A$ & Common Words & tfidf$_R$ & tfidf$_A$ & tfidf$_R - $tfidf$_A$ \\ \hline
posted & 0.1233 & 0.0002 & 0.1230 & alpha & 0.0830 & 0.4820 & -0.3990 \\ \hline
\textgreek{βιβλίο}, & 0.1182 & 0.0003 & 0.1179 & \textgreek{αγόρασέ} & 0.1358 & 0.0809 & 0.0549 \\ \hline
\textgreek{ίδρυμα} & 0.0906 & 0.0007 & 0.0899 & ekdromi.gr & 0.1258 & 0.0624 & 0.0634 \\ \hline
\textgreek{κωδικός} & 0.0830 & 0.0079 & 0.0751 & hotel & 0.0704 & 0.0608 & 0.0096 \\ \hline
\textgreek{τράπεζας} & 0.0830 & 0.0001 & 0.0829 & newsletter & 0.0453 & 0.0560 & -0.0107 \\ \hline
\textgreek{όνομα}, & 0.0830 & 0.0006 & 0.0825 & \textgreek{εικόνα} & 0.0629 & 0.0483 & 0.0146 \\ \hline
\textgreek{γιάννης} & 0.0805 & 0.0014 & 0.0791 & \textgreek{έκδοση} & 0.0528 & 0.0470 & 0.0058 \\ \hline
subscribed & 0.0780 & 0.0013 & 0.0767 & \textgreek{διαθέσιμη} & 0.0478 & 0.0454 & 0.0024 \\ \hline
states & 0.0755 & 0.0001 & 0.0754 & column & 0.0453 & 0.0392 & 0.0061 \\ \hline
united & 0.0755 & 0.0001 & 0.0753 & outlook & 0.0428 & 0.0322 & 0.0106 \\ \hline
\end{tabular}
\end{table*}

\begin{table*}[htbp]
\centering
\caption{TF-IDF results for the Romanian language variant.}
\label{tab:tfidf-romanian}
\begin{tabular}{|l|l|l|l|l|l|l|l|}
\hline
\multicolumn{1}{|c|}{Searched Words} & \multicolumn{1}{c|}{tfidf$_R$} & \multicolumn{1}{c|}{tfidf$_A$} & \multicolumn{1}{c|}{tfidf$_R - $tfidf$_A$} & \multicolumn{1}{c|}{Common Words} & \multicolumn{1}{c|}{tfidf$_R$} & \multicolumn{1}{c|}{tfidf$_A$} & \multicolumn{1}{c|}{tfidf$_R - $tfidf$_A$} \\ \hline
posted & 0.2307 & 0.0011 & 0.2296 & click & 0.1567 & 0.2693 & -0.1127 \\ \hline
charm & 0.1481 & 0.0038 & 0.1443 & multe & 0.1253 & 0.2238 & -0.0984 \\ \hline
dimensiune & 0.1424 & 0.0024 & 0.1401 & \'E\textsuperscript{TM}te & 0.0541 & 0.1470 & -0.0928 \\ \hline
greutate & 0.1424 & 0.0045 & 0.1379 & adresa & 0.0741 & 0.1436 & -0.0696 \\ \hline
numar & 0.1339 & 0.0093 & 0.1245 & romania & 0.0427 & 0.1161 & -0.0734 \\ \hline
cutiuta & 0.1253 & 0.0017 & 0.1237 & online & 0.0627 & 0.1118 & -0.0491 \\ \hline
livreaza & 0.1253 & 0.0019 & 0.1234 & video & 0.0968 & 0.1085 & -0.0117 \\ \hline
argint & 0.1310 & 0.0103 & 0.1207 & dintre & 0.0826 & 0.1037 & -0.0211 \\ \hline
material & 0.1253 & 0.0068 & 0.1185 & dezabonare & 0.0370 & 0.0992 & -0.0622 \\ \hline
produsul & 0.1253 & 0.0089 & 0.1164 & iulie & 0.0826 & 0.0991 & -0.0165 \\ \hline
\end{tabular}
\end{table*}

Tables \ref{tab:tfidf-english}, \ref{tab:tfidf-greek}, and \ref{tab:tfidf-romanian} show the results of TF-IDF analysis on English, Greek, and Romanian honey accounts respectively. They show that those who accessed the Greek and Romanian accounts attempted to search for words outside the linguistic confines of the accounts. For instance, the word ``posted'' appeared to be the most searched word in both the Greek and Romanian accounts. The terms searched in the Romanian accounts did not include any financial or banking indicators, whereas the TF-IDF search approximation for the Greek accounts includes words such as \textgreek{τράπεζας} (bank) and \textgreek{κωδικός} (code). Both words are among the sensitive terms that we used to seed the accounts beforehand, as earlier described in Section \ref{sec:methodology:creating}. On a related note, financial terms such as ``banking'' and ``investment'' appear among the top TF-IDF words in the English accounts (see Table \ref{tab:tfidf-english}). These findings show that cybercriminals indeed searched for financial terms in the honey accounts. This result is further strengthened by the observation that the terms found to be important in the entire email text $d_A$ are not important in the corpus of opened emails $d_R$ (as shown by the low tfidf$_R - $tfidf$_A$ values, some of which are negative). This is a strong indicator that the opened emails were not selected randomly by the cybercriminals, rather, they were opened deliberately after searches were conducted for those terms. This further corroborates findings from previous work~\cite{bursztein2014handcrafted, onaolapo2016gmail}.

\section{Discussion} 
\label{sec:discussion}

In this section, we provide a summary of our findings and the limitations of our approach. Finally, we discuss potential future work.  

\desc{Summary of our findings}Contrary to our expectations, our findings show that cybercriminals are more likely to locate sensitive information in the Greek accounts than accounts in the other languages. This is rather intriguing, especially since only two of the accesses we observed originated from Greece or Greek-speaking countries. We recognize that some accesses to the accounts may have been made through proxy servers. However, it is clear that those who visited the accounts were not solely Greek-speaking individuals. These findings run contrary to the ideas espoused in theories of language comprehension and understanding, which suggest that individuals should be significantly hindered in their comprehension if they do not understand the language of the object they are interacting with. Thus, we postulate that the cybercriminals possibly used online language translation tools to translate financial terms to Greek prior to searching the Greek accounts for such keywords. This would also explain the amount of time that they spent accessing the accounts: Greek accounts recorded longer access times than the rest, while English accounts recorded the lowest. 

Miscreants spend more time on average going through the Greek and Romanian accounts. This indicates a number of possibilities. As earlier stated, cybercriminals may spend more time on the accounts to incorporate the use of online translation services to improve their limited understanding of email content, thus spending more time on those accounts. Alternatively, it may be because individuals are more readily able to assess the contents of a webmail account whose content language is English, and consequently disregard such an account if it appears to be of limited value. 

Finally, the implementation of a way to search for keywords in the content of an email account may be a key factor in the ability of a criminal to traverse a compromised webmail account, as seen in our TF-IDF evaluation which highlighted words such as ``bank'' and ``code.'' This suggests that it might be possible for webmail service providers to hamper criminal elements from finding sensitive information in compromised accounts by obfuscating or removing keywords relating to banking or financial matters.    

\desc{Limitations} First, we were able to leak the honey accounts through paste sites only. Hence, our results may not reflect what happens to accounts that are compromised via other outlets. Second, our approach relies on TF-IDF to approximate search terms in the honey accounts. As a result, we only have insight into searches whose results were opened by the miscreants. We are unable to assess searches that did not return results, and searches that returned results which the miscreants did not open.    

\desc{Future work} In the future, we intend to explore the use of compromised online accounts in other scenarios, for instance, in targeted attacks. We also intend to study the impact of language differentiation on cybercriminal activity on other platforms, for instance online social networks, cloud storage accounts, and online banking accounts.

\section{Related Work} \label{sec:related}
  
Bursztein et al.~\cite{bursztein2014handcrafted} studied the use of compromised Gmail accounts in the wild, with specific focus on spearphishing as a way by which cybercriminals obtain account credentials. They deployed Gmail honeypots and collected data from them. In our previous paper~\cite{onaolapo2016gmail}, we used a similar honeypot approach to investigate the use of compromised Gmail accounts, but explored more outlets, namely paste sites, underground forums, and malware. We also presented a public honeypot infrastructure, which we used in this paper. Other researchers have used honeypot systems to study the use of compromised online accounts as well. Liu et al.~\cite{liu2012many} placed honey credentials (inside honey files) in P2P shared spaces to study illegitimate accesses. Nikiforakis et al.~\cite{nikiforakis2011exposing} also studied privacy issues in file hosting systems using honeyfiles. Stringhini et al.~\cite{Stringhini:10:socialnet-spam} deployed honeypot profiles to study social spam. Other studies exploring the misuse of online accounts include~\cite{Benvenuto:2010:CEAS, boshmaf2011socialbot, lee2010social, thomas2011suspended}. They focus on the abuse of online accounts, while we focus the effect of language differentiation on the ability of cybercriminals that attempt to abuse webmail accounts and steal sensitive information. 

\section{Conclusion}

In this paper, we studied the impact of language differentiation on the activity of cybercriminals accessing compromised webmail accounts. We created, deployed, and leaked thirty honey accounts across three languages, namely English, Greek, and Romanian. We collected and analysed data about accesses and activity from the honey accounts for \durationOfExpt. Our tests revealed a significant relationship between language and the ability of a cybercriminal to access a sensitive item (that we seeded the account with). Finally, we presented the results of our analysis on the contents of the honey accounts. We found that cybercriminals indeed searched for sensitive financial information in the accounts. We hope our findings will help the research community to gain deeper insight into the relationship between language and cybercriminal activity, and potentially provide insight into ways to develop effective techniques to detect illegitimate activity in online accounts. 

\bibliographystyle{IEEEtran}
\bibliography{biblio}

\end{document}

%% file: vars.tex
\newcommand\numOfAccounts{thirty\xspace}
\newcommand\durationOfExpt{fifteen days\xspace}
\newcommand\numOfAccesses{650\xspace}
\newcommand\numOfCountries{nineteen\xspace}